\documentstyle[11pt,paspconf,epsf]{article}

\begin{document}

\title{Discovery of a New Filamentary Structure G358.85+0.47 - The Pelican}
\author{K.R. Anantharamaiah$^{1,2}$, C.C. Lang$^{1,3}$, N.E. Kassim$^4$
        T.J.W. Lazio$^4$ and W.M. Goss$^1$}
\affil{$^1$National Radio Astronomy Observatory, Socorro, NM 87801, USA}
\affil{$^2$Raman Research Institute, Bangalore 560 080, India}
\affil{$^3$Dept of Astronomy, UCLA, Los Angeles, CA, USA}
\affil{$^4$Naval Research Laboratory, Washington D.C., USA}

\index{Source!G358.85+0.47}
\index{Source!Pelican}
\index{Galactic Center!Filaments}

\begin{abstract}

We report the discovery of a new filamentary structure, G358.85+0.47,
consisting of at least three mutually parallel but bent `strands',
located about 1.5$^\circ$ SW of the Sgr A complex. Unlike all the
other known Galactic center filaments, G358.85+0.47 is oriented
parallel to the galactic plane. Based on its appearance in a 20 cm
image, we give it the name 'the Pelican' for further reference.

\end{abstract}

\section{Introduction}

The region within about 150 pc of the Galactic Center contains many
unique radio structures. Particularly interesting among these are
several linear, non-thermal structures termed "arcs", "threads", and
"filaments". The Radio Arc which is a bundle of about a dozen parallel
filamentary structures was discovered by Yusef-Zadeh, Morris and
Chance (1984). Two isolated linear structures known as "threads"
(G0.08+0.15 and G359.96+0.09) were first reported by Morris and
Yusef-Zadeh (1985). The other known linear structures near the
Galactic center are G359.79+0.17, Sgr C (Liszt 1985), G359.54+0.18
(Bally and Yusef-Zadeh 1989) and G359.1-0.2 or the "snake" (Gray et al
1991). The linear structures are tens of parsecs long and only a
fraction of a parsec wide. Their origin is still not understood. Some
of the known properties of these filaments are (Morris 1996, Lang
1999): (1) most of them seem to be oriented perpendicular to the
Galactic plane and lie at positive Galactic latitudes,  (2) the radio
emission is non-thermal and strongly linearly polarized, (3) the
spectra of the filaments, where they have been measured, is steep
($S\propto\nu^{-0.6}$), or, flat, or even inverted  (Anantharamaiah
et al 1991),  (4) there are no obvious radio point sources associated
with the filaments and (5) in every well studied case, there appears to
be an association with a molecular cloud.

Although the origin of these filaments is still enigmatic, there is
general agreement that there is an intimate connection between these
filaments and the large scale magnetic field structure in the Galactic
center region.  In fact, most of the models that are proposed to
explain these structures, make use of some type of magnetic phenomenon
including generation of magnetic loops (Heyvaerts, Norman and Pudritz
1988), induced electric field through the {\bf $v\times B$}
interaction between a moving molecular cloud and a uniform field
(Benford 1988), particle acceleration by field line annihilation
around current pinches (Lesch and Reich 1992) and field annihilation
at the surface of molecular clumps (Serabyn and Morris 1996).
 
In most of the models, the filaments are thus thought to define the
magnetic field direction. Polarization measurements of some of the
filaments do support this idea (Lang 1999). As these threads and
filaments are oriented essentially perpendicular to the Galactic
plane, these radio structures are often considered as strong evidence
for the presence of an intense dipolar poloidal magnetic field in the
galactic center region (Morris 1996).  

In this paper we report the discovery of a new filamentary structure
which is about 1.5$^\circ$ SW of the Sgr~A complex. This filamentary
structure is unique in that it is oriented parallel to the galactic
plane whereas all the other previously known filaments are oriented
roughly perpendicular to the galactic plane. The orientation of this
filament may have implications for the large scale structure of the 
magnetic filed in the galactic center region.

\begin{figure}[h]
\plotfiddle{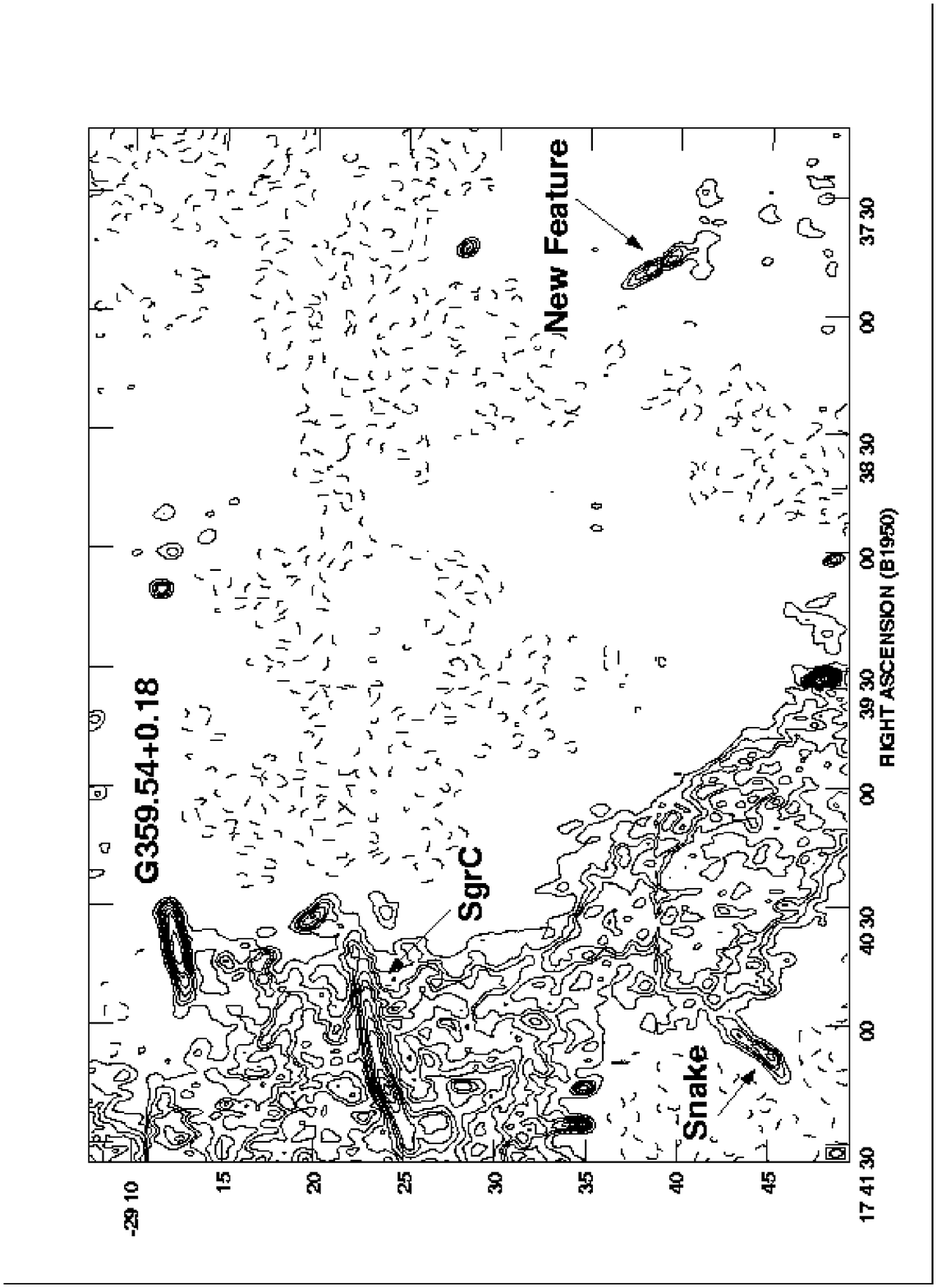}{2.6in}{-90}{38}{38}{-170}{+220}
%\plotone{gcfig1.ps}
%\vspace{2.5 in}
\caption{A sub-region of the wide-field image at 90 cm showing the
new filament and three other previously known filaments. The beam
depicted at the bottom left corner is $38''\times 25''$. The contour
levels are -10,10,20,30,40,60,80,100, 140, ... 620 mJy/beam}
\end{figure}

\section{Identification of the New Filament in a Wide-field 90 cm Image}

Using the data at $\lambda$ = 90 cm from all the four VLA
configurations obtained by Pedlar et al (1989) and Anantharamaiah et
al (1991), a new wide-field ($\sim 4^{\circ}$) radio image of the
Galactic center region was recently made by Kassim et al (1999). Fig 1
shows a sub-region of this image.  The new image was made using a 3D-imaging
algorithm developed by Cornwell and Perley (1992) which corrects for
the non-coplanarity of the VLA and allows sensitive, distortion-free
imaging of sources far from the field center. In a careful examination
of the image made using the B+C+D configuration data with a resolution
of $38''\times 25''$, we identified a new linear structure which is
located about 1.5\deg south-west of SgrA. In Fig 1, the new linear
feature is located in the bottom right corner.  The band of emission
on the left hand side of Fig 1 is the emission from the Galactic
plane. Fig 1 also shows three other previously known filamentary
structures G359.54+0.18, Sgr~C and the G359.1-0.2 (the Snake).  The
new linear feature is located at the position $\alpha_{1950} = 
17^h37^m47^s.6$ and $\delta_{1950} = -29^\circ 38'36''$
and is similar in appearance to the other linear
structures but is oriented parallel to the Galactic plane. We
designate this feature as G358.85+0.47 based on its galactic
coordinates.

\begin{figure}[h]
\plotfiddle{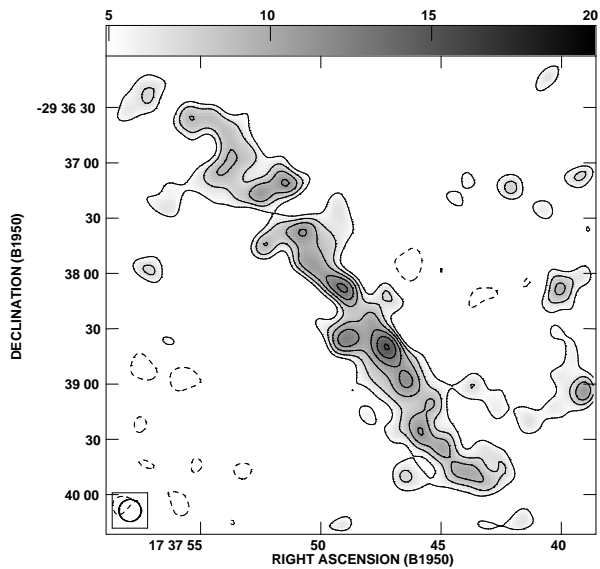}{2.7in}{0}{40}{40}{-400}{-20}
%\vspace{2.5 in}
\caption{Higher resolution ($12''$) image of G358.85+0.47 at 90 cm
made using only the A\&B configuration data. The contour levels are -5,5,
7,9,11,13,15,17 mJy/beam}
\end{figure}

This newly found linear feature was examined further at higher
angular resolution using an image made with only the A and B
configuration data.  Fig 2 shows a 90 cm image of G358.85+0.47 at a
resolution of $12''$. G358.85+0.47 has a linear extent of $\sim 4'$
and appears barely resolved along its width. There are a few spots of
enhanced emission along its length. The brightness of  G358.85+0.47
ranges from 10 to 15 mJy/beam at $\lambda$=90 cm.

\section{Follow up Observations at $\lambda$ = 20 cm}

Follow up observations of G358.85+0.47 were made on Aug 24, 1998 using
the B-configuration of the VLA at $\lambda$ = 20 cm. Data was acquired
in two frequency bands each of width 50 MHz and two orthogonal (RR and
LL) polarizations. The amplitudes were calibrated using the source
3C286 and initial phase calibration was performed using frequent
observations of the source 1748-253.  The final image made after
self-calibration and combining both the frequency bands is shown
in Fig 3 in grey scale and in Fig 4 in contours. 

\begin{figure}[h]
\plotfiddle{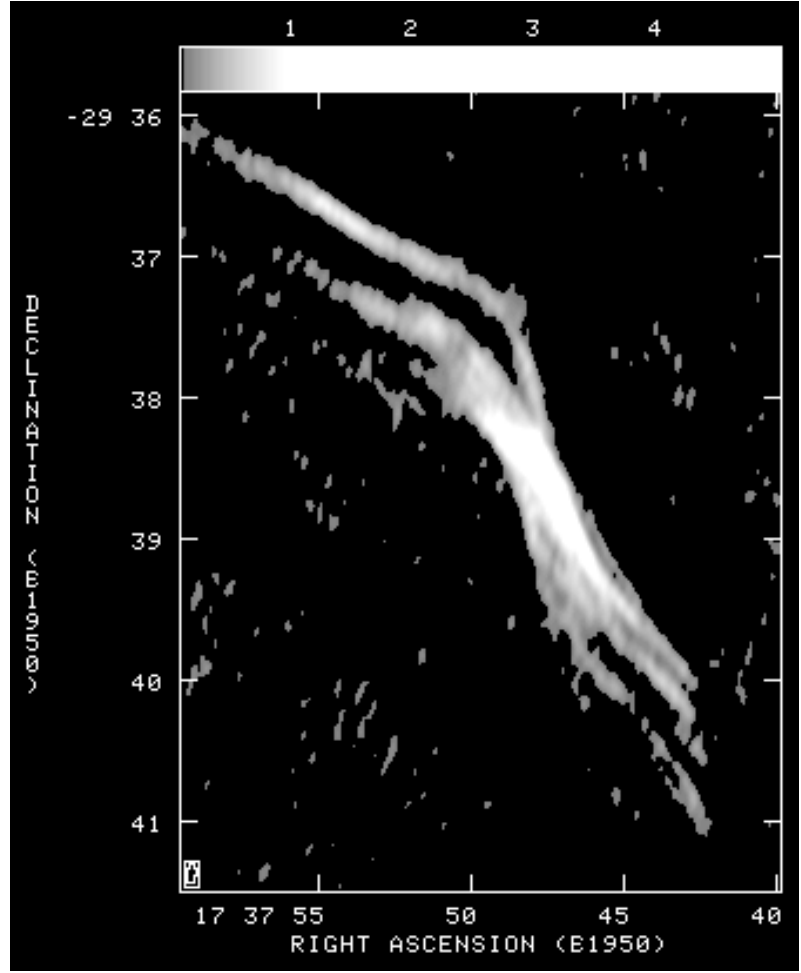}{5.0 in}{0}{55}{55}{-180}{-30}
%\vspace{2.5 in}
\caption{20 cm image of the new filament  G358.85+0.47 in Gray scale.
The beam size is $8.4'' \times 3.4''$, PA = -7$^\circ$.}
\end{figure}

The filamentary nature of the new feature is obvious from Figs 3 and 4. There
could be little doubt that this feature belongs to the same category
of radio filaments such as Sgr~C, G359.54+0.18, G359.79+0.17, the
Snake and the 'threads' G0.08+0.15 and G359.96+0.09. The
distinguishing feature of the new filament is that it is oriented
parallel to the galactic plane. Although we do not yet have any radio
polarization data for the new filament, based on what is known about
the other filaments (Lang 1999), it is reasonable to expect that the
magnetic field lines will be aligned along the length of the filament
or in other words parallel to the galactic plane. Such a orientation
of the field lines will have to be reconciled with the strong poloidal
magnetic field that has been postulated based on the orientation of the
previously known filaments. It is possible that the location and
the orientation of the new filament may set limits on the distance
from the galactic center up to which a dipolar field may be extended.
For example, the new filament may indicate the position where the
field lines are turning around to close the loop.

\begin{figure}[h]
\plotfiddle{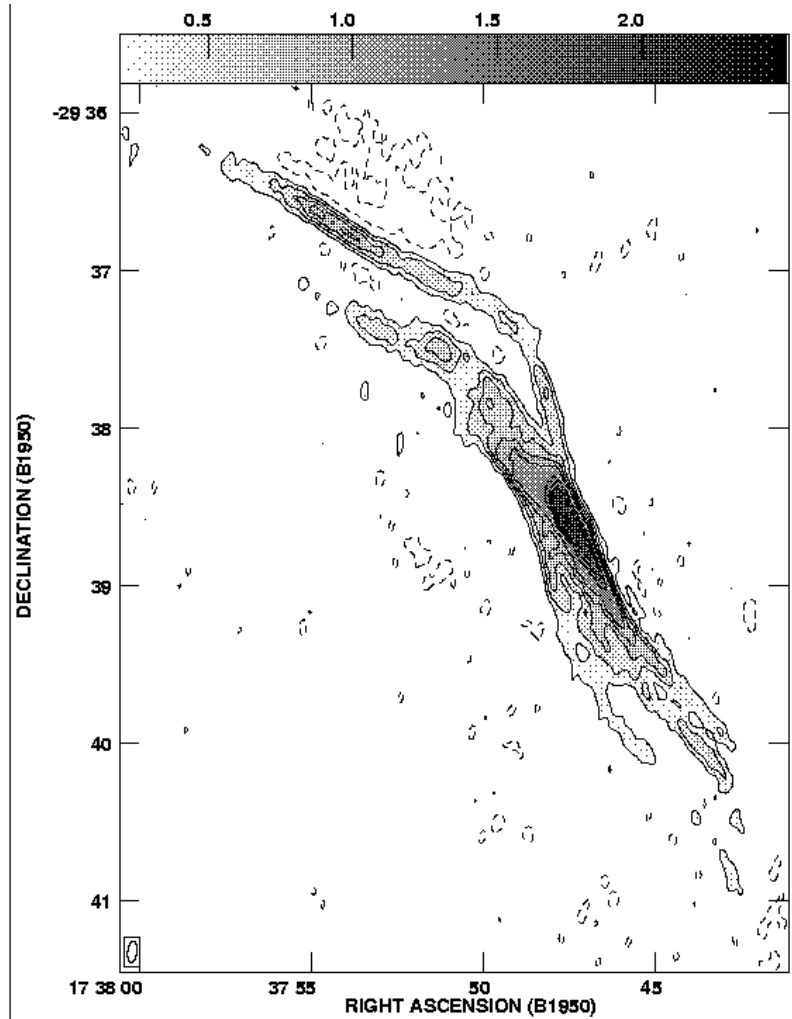}{4.9 in}{0}{53}{53}{-190}{-30}
%\vspace{2.5 in}
\caption{20 cm image of the new filament  G358.85+0.47 in contours.
Contour levels are -0.2,0.2,0.4, 0.6 ... 2.2 mJy/beam. The beam is
depicted at the bottom left corner ($8.4''\times 3.4''$, PA = -7$^\circ$)}
\end{figure}

The overall structure of the new filament is also unique. It 
consist of at least three sub-filaments which are bent in an inverted
S-shape.  The bend in the northern most filament is particularly
pronounced. The distortion in the filaments implies either that the field
lines in this regions are themselves distorted or that the structure
of the filament is not dominated by the magnetic field.  The minimum
magnetic field in the northern sub-filament implied by assuming that
radio emission at 1.4 GHz is due to synchrotron emission is 70 $\mu G$.

It is not clear why there are significant differences in the structure
of the filament at 90 cm and 20 cm (Figs 2 and 4). The differences
persist even when the images are compared at the same angular
resolution ($12''$). While the filamentary structure is obvious in the
20 cm image, at 90 cm there are several peaks along the length of the
filament and the northern most bent filament is absent in Fig 2.  The
90 cm image is limited by the dynamic range of the wide-field image.
Note that at 90 cm, the field of view of the VLA includes the emission
from the entire Sgr~A complex and thus the image in Fig 2 had to be
made after subtracting strong emission from other regions. Because of
these differences, the determination of the spectral index between 90
cm and 20 cm is uncertain. At the peak of the source the derived
spectral index is $-0.1\pm 0.2$. While this spectral index is
consistent with the flat or inverted spectrum exhibited by several
other filaments (Anantharamaiah et al 1991), higher frequency
observations are required for a better estimate.

As there are a number of unique radio features near the galactic
center, it is essential to assign identifying names to these features
for easy reference. Based on the visual appearance of its image in Fig
3, we suggest the name 'Pelican' for the new filament.  Further
observations of the pelican-shaped filament in Fig 3 are underway to
determine its radio spectral index, polarization and possible
association with molecular gas.

The National Radio Astronomy Observatory is a facility of the National
Science Foundation, operated under a cooperative agreement by Associated
Universities, Inc.

\end{document}